\documentclass[pre,showpacs,twocolumn]{revtex4-1}
\usepackage{graphicx}
\usepackage{amsmath}
\usepackage{amsfonts}
\usepackage{hyperref}
\usepackage{url}
\usepackage{color}
\usepackage{array} 
\usepackage{mathtools}
\usepackage{multirow}
\renewcommand{\AA}{\ensuremath{\mathbb{A}}}
\newcommand{\GG}{\ensuremath{\mathbb{G}}}
\newcommand{\TT}{\ensuremath{\mathbb{T}}}
\newcommand{\ttt}{\ensuremath{t \kern -2pt t}}

\usepackage{Tomspeak}
\newcommand{\tw}[1]{{\color{blue}[\kern-2pt [\hbox{\lower.5ex\hbox{\tiny TW}} #1]\kern -2pt]}}
\newcommand{\am}[1]{{\color{red}[\kern-2pt [\hbox{\lower.5ex\hbox{\tiny AM}} #1]\kern -2pt]}}

\renewcommand{\vector}[1]{\ensuremath{\boldsymbol{#1}}}
\newcommand{\crossmat}[1]{[\kern-2pt [ #1 ^\times] \kern -2pt]}
\newcommand{\cf}{\textit{cf.} }
\begin{document}

\author{Aaron J. Mowitz}
\email{amowitz@uchicago.edu}
\author{T. A. Witten}
\email{t-witten@uchicago.edu}
\affiliation{Department of Physics and James Franck Institute, University of Chicago, Chicago, Illinois 60637, USA} 
\title{Predicting tensorial electrophoretic effects in asymmetric colloids}
\date{\today}

\begin{abstract}

We formulate a numerical method for predicting the tensorial linear response of a rigid, asymmetrically charged body to an applied electric field.  This prediction requires calculating the response of the fluid to the Stokes drag forces on the moving body and on the countercharges near its surface.  To determine the fluid's motion, we represent both the body and the countercharges using many point sources of drag known as stokeslets.  Finding the correct flow field amounts to finding the set of drag forces on the stokeslets that is consistent with the relative velocities experienced by each stokeslet.  The method rigorously satisfies the condition that the object moves with no transfer of momentum to the fluid.  We demonstrate that a sphere represented by 1999 well-separated stokeslets on its surface produces flow and drag force like a solid sphere to one-percent accuracy.  We show that a uniformly-charged sphere with 3998 body and countercharge stokeslets obeys the Smoluchowski prediction \cite{Morrison} for electrophoretic mobility when the countercharges lie close to the sphere.  Spheres with dipolar and quadrupolar charge distributions rotate and translate as predicted analytically to four percent accuracy or better.  We describe how the method can treat general asymmetric shapes and charge distributions.  This method offers promise as a way to characterize and manipulate asymmetrically charged colloid-scale objects from biology (\eg viruses) and technology (\eg self-assembled clusters).
\end{abstract}
\maketitle

\section{Introduction}
\label{sec:Introduction}
\begin{figure}
\includegraphics[width=.47\textwidth]{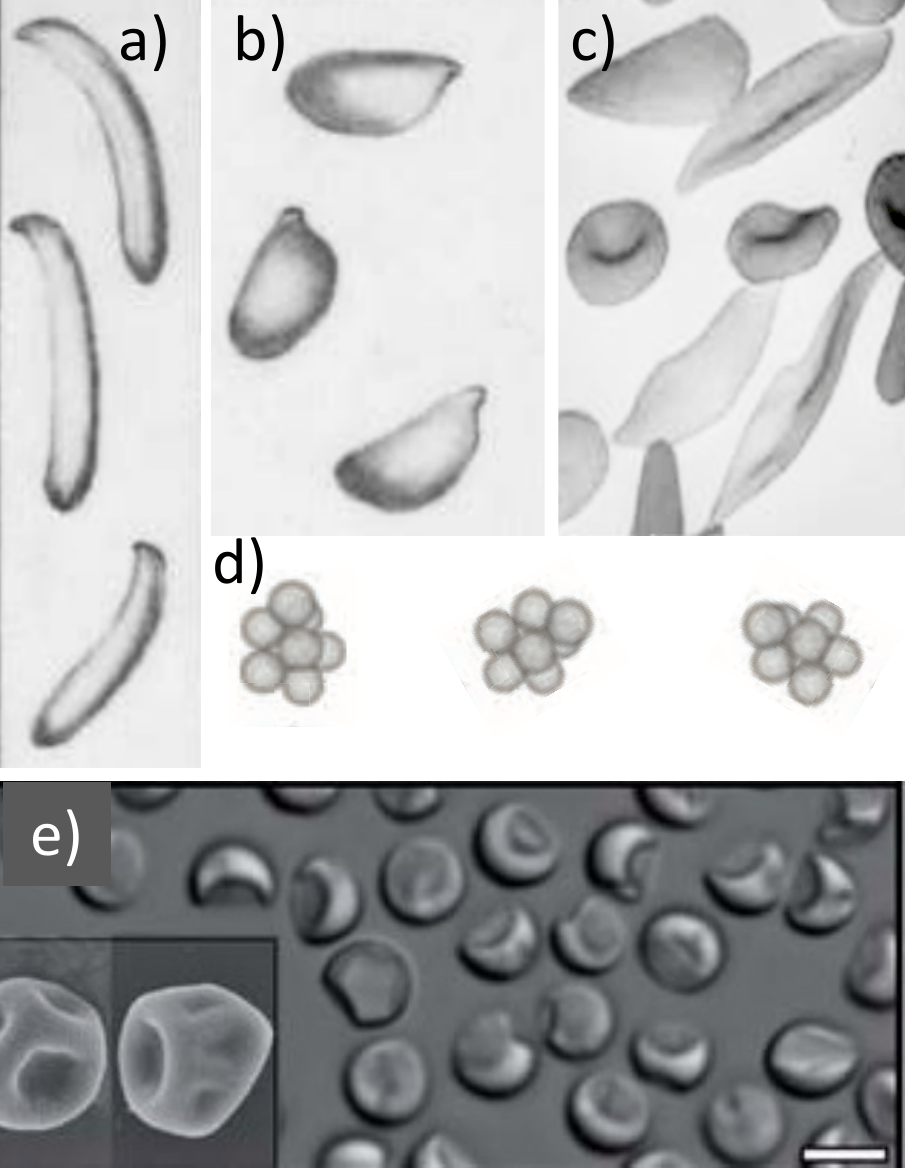}
\caption{\label{fig:objects}
Asymmetrical, self-assembled micron-scale objects from nature and technology. 
a) Spores of Tilletsiopsis fungi after Fig. IX of \cite{DERX:1930fk}. Long dimension is 8-10 micron
b) Spores of Sporobolomyces odorus fungus after Fig. III of \cite{DERX:1930fk}.  Long dimension is about 7 micron.
c) Human red blood cells, after \cite{Jayawickrama:2013li}.  Normal disc-shaped cells in center have a diameter of about 7 microns.  Elongated cells are deformed by hemoglobin aggregates.
d) Most-probable eight-sphere colloidal cluster assembled from 1-micron polystyrene spheres by depletion interaction, after \cite{Meng560},  Fig. 2  .
e) Collapsed colloidal shells after \cite{Sacanna:2011fk}.  Scale bar is 3 microns.
}
\end{figure}

Biology provides us many examples where the asymmetric shape and spatial arrangement of a micron-scale object enables its distinctive function.  Often these objects, such as a virus, a bacterium, a red blood cell or yeast cell have these conserved spatial features, as Fig. \ref{fig:objects} illustrates \cite{lodish2008molecular}.  It is now also possible to create man-made colloidal objects with reproducible, distinctive shapes via self assembly methods \cite{Meng560,Sacanna:2011fk}.  Here too the distinctive spatial structure produces distinctive motions and interactions within the fluid environment of these objects.

One consequence of asymmetric shape is a complex, tensorial response to an electric field; these responses have been anticipated \cite{AjdariLong} but have not been well explored.     The objects translate and rotate in directions that depend on their orientation relative to the field.  This response gives a means of probing and manipulating these objects. For example, measuring the tensorial responses of a virus would specifically reveal asymmetries in its charge distribution, complementing other structural probes.  Once the response tensors have been determined for a given object, it can be steered and manipulated via suitable time-varying fields \cite{Moths-Witten1} \eg to separate objects with different charge distributions.

Here we demonstrate a discrete-particle numerical method to predict electric-field-induced motion of asymmetric shapes straightforwardly.  Comparison with known results for asymmetrically charged spheres demonstrates the feasibility and accuracy of the method.  This method enables exploration of different objects not feasible hitherto.  

The driven motion of colloidal particles in a fluid is a classical and well-studied subject \cite{Happel-Brenner}.  Topics of high current interest include active agents that produce internal driving \cite{Marchetti:2013pi,Zottl:2014ye,Aranson:2013kx}, non-newtonian fluids \cite{Akers:2006lk} and hydrodynamically-interacting suspensions \cite{TomerHaim1,TomerHaim2}.  Here we focus on the distinctive responses owing to the geometric features of the colloidal object.  The simplest case of such a response is sedimentation, in which an external body force $\vector F$ acts at a certain point in the object.   The resulting motion includes translation, with velocity $\vector V$ and rotation with angular velocity $\vector \Omega$.   The strongest predictive power and experimental relevance for these motions lies in the regime of linear response, where $\vector V$ and $\vector \Omega$  are proportional to the force $\vector F$ via tensors defined with respect to the body's orientation.  Because of these tensors, the body may execute cyclic, chiral motions even in steady state.  If $\vector F$ varies in time, the responses are richer and can be used to organize a dispersion of many identical objects \cite{Eaton-Moths-Witten}.  These responses are of potential value for characterizing and organizing objects according to their shape.  However, the practical value for colloid-scale objects is limited because the response is weak.  

The situation is more favorable for electrophoresis, motion induced by a uniform electric field $\vector E_0$ \cite{Lyklema}. As with sedimentation there is a broad, experimentally relevant regime in which the response is linear in $\vector E_0$ and is completely governed by two tensors $\AA$ and $\TT$ defined by $\vector V = \AA \vector E_0$ and $\vector \Omega = \TT \vector E_0$.  These tensors depend on the object's shape and charge distribution.  They also depend on the electrical screening properties of the surrounding fluid.  We focus on the regime of strong linear screening \cite{*[{}] [{ Chap. 14}] Israelachvili:2011qc} attained \eg in ordinary salty water.   This regime gives the greatest predictive power and experimental accessibility.  We shall consider objects with small enough charge density that the screening response is described by the linear Debye theory \cite{Israelachvili:2011qc}.  The electrostatic screening length $\lambda_D$ in these conditions is typically a few Angstroms; we shall consider only objects that are much larger than $\lambda_D$.  

Electrophoretic motion through a fluid is qualitatively different from that of sedimentation.  The sedimentation response is dominated by the momentum transmitted to the fluid by the driving force $\vector F$.  In electrophoresis, {\it no} net momentum is transmitted to the fluid; all the force exerted by the object is canceled by the neutralizing screening charge \cite{Israelachvili:2011qc}.  Instead, the object crawls through the fluid by displacing a thin sheath of fluid of thickness $~\lambda_D$ around the object.   This makes possible a broader range of anisotropic responses than that seen in sedimentation.  As Long and Ajdari demonstrated \cite{AjdariLong}, the object may rotate indefinitely without translating or may translate indefinitely in a direction perpendicular to any applied $\vector E_0$ field.  

The Long-Ajdari examples suggest a wide range of possibilities for real colloidal objects, though their calculations apply to objects that are difficult to realize experimentally.  Therein lies the motivation for the present work.  We seek to provide a practical means to infer the mobility tensors $\AA$ and $\TT$ for realistic colloidal objects of a given shape and charge distribution.  Procedures for calculating these tensors have long been known \cite{Teubner,AndersonSpheres} and have been applied for simple shapes such as spheres and ellipsoids \cite{AndersonSpheres,AndersonEllipsoids}.  One must first determine the screening charge distribution around the object.  Next one must determine the local electric field around the object in the presence of the applied field $\vector E_0$.  Then one may determine the external force acting on an element of the fluid with nonzero charge density under the local electric field.  Finally one may calculate the velocity profile of the fluid owing to these external forces, as modified by the no-flow boundary condition at the nearby body surface.   This velocity profile dictates the motion of the object and thence $\vector V$ and $\vector \Omega$.  Though straightforward in principle, this prescription is formidable in practice, since each step requires a separate numerical determination of a three-dimensional field under asymmetric boundary conditions. 

We aim to circumvent these complexities using a simplified approximate representation of the object and its charge distribution. The approximate object is chosen to allow an explicit determination of the fluid motion. Specifically, we represent the object as a set of point sources of fluid drag called stokeslets \cite{Kirkwood-Riseman}.  Representing neutral bodies in this way is a well-established procedure \cite{Chen:1987jo}.  In Section \ref{sec:stokeslets} below we show how this procedure is used for sedimentation, noting that even well-separated stokeslets can represent a solid body. Next, in Section \ref{sec:charged} we describe the basic features of electrophoresis. Section \ref{sec:electrophoresis} describes the additional ingredients needed in our procedure in order to treat electrophoresis: the distortion of the electric field by the body and the forces on the fluid from the screening charge.  In Section \ref{sec:spheres} we give an explicit prescription for spherical bodies, where we may test our methodology against known results.  In Section \ref{sec:results} we describe our numerical results and compare them to the established results.  In Section \ref{sec:discussion} we discuss limitations and generalizations of our method.

\section{Stokeslet objects}\label{sec:stokeslets}

In this section we discuss the hydrodynamic representation of an asymmetric object as an assembly of stokeslets.  Such ``stokeslet objects" have the advantage that the Stokes-law flow field around them can be calculated explicitly by a numerical matrix inversion.  

The flow created by a single stokeslet is well known.  If a force $\vector f$ is exerted on a quiescent fluid of viscosity $\eta$, it gives rise to a steady-state velocity field $\vector v(\vector r)$ at a displacement $\vector r$ from the forced point.  The velocity is proportional to $\vector f$ via the ``Oseen tensor" $\mathbb{G}({\vector r})$ \cite{Happel-Brenner}:
\begin{equation}
  \vector v = \mathbb{G}({\bf r})\cdot\vector f, \hbox{where} \ \ \
  \mathbb{G}_{ij}({\vector r}) = \frac{1}{8\pi \eta r} \left( \delta_{ij} + 
  \frac{r_ir_j}{r^2} \right).
\label{eq:Oseen}
\end{equation}
An assembly of $N$ such sources $\vector f^{\beta}$ located at $\vector r^\beta$ produces a velocity at any given point  $\vector r^\alpha$ given by superposition: 
\begin{equation}
\vector v(\vector r^{\alpha}) = \sum_{\beta \notequal \alpha} \GG(\vector r^\alpha - \vector r^{\beta}) ~ \vector f^\beta
\label{eq:StokesletVelocity}
\end{equation}

If there is a particle at $\vector r^\alpha$, it transmits to the fluid a drag force $\vector f^\alpha$ opposing the fluid's motion relative to the particle. We may suppose the particle to be a tiny sphere of radius $a$, whose drag coefficient $\Gamma$ is then $6\pi\eta a$ \cite{Brady:1988uq}.  Then if the velocity of the particle is $\vector U^\alpha$, the force $\vector f^\alpha = -\Gamma~(\vector v(\vector r^\alpha) - \vector U^\alpha$).  
We now suppose all the sources in Eq. \eqref{eq:StokesletVelocity} to be similar spheres, which are constrained to move as a rigid body with a common velocity $\vector V$ and angular velocity $\vector \Omega$ relative to some chosen origin in the body.  Thus $\vector U^\alpha = \vector V + \vector \Omega \times \vector r^\alpha$.  Using these constraints in Eq. \eqref{eq:StokesletVelocity} yields 
\begin{equation}
\vector f^\alpha = -\Gamma \left (\sum_{\beta \notequal \alpha} \GG(\vector r^\alpha - \vector r^{\beta}) ~ \vector f^\beta - \vector V - \vector \Omega \times \vector r^\alpha \right )
\label{eq:fConsistency}
\end{equation}
This linear system can readily be solved for the unknown $\vector f$'s.  One may sum these forces to obtain the total drag force and torque on the body in terms of the given $\vector V$ and $\vector \Omega$.  

This stokeslet prescription was devised \cite{Kirkwood-Riseman} to represent tenuous objects like random-coil polymers \cite{Gennes:1979qd} or colloidal aggregates \cite{Chen:1987jo}, where the stokeslets are placed at each monomer or subunit of the object.  Nevertheless it is adequate to represent solid objects such as spheres, as illustrated in Fig. \ref{fig:sedelph}.  Dilute stokeslets can represent a solid object because of hydrodynamic screening, as discussed in Appendix \ref{sec:screening}. There we derive a nominal penetration depth $\xi_e$ in terms of the area fraction $\phi_s$ of stokeslets and their radius $a$:
\begin{equation}
\xi_e \definedas a/\phi_s
\end{equation}  
Thus for any area fraction $\phi_s$ however small, one can attain a desired small $\xi_e$ by choosing small enough spheres \cite{Chen:1987jo}. Fig. \ref{fig:sedelph} shows that a monolayer of 1999 stokeslets faithfully represents the flow around a solid sphere.  The flow near the sphere surface is smoothed at a scale which here is evidently much smaller than the sphere radius.  Similar accuracy is expected for smooth objects of generic form.

We now consider the effect of electric forces on the object.
\section{				Charged body in electric field}
\label{sec:charged}
\subsection{			Electrically insulating bodies and depolarization field }\label{sec:dpfield}
In addition to blocking fluid flow, an electrically insulating body in a conducting medium must exclude any current from its interior.  Thus the current just outside the surface of the body must be tangent to that surface.  Since this current is proportional to the electric field, the surface electric field $\vector E_s$ must be tangent to the surface as well \cite{Morrison}.  Thus when a uniform exterior field $\vector E_0$ is applied, the blockage of current from the interior gives rise to a depolarization charge $\rho_d$.  The normal electric field from this charge is just such as to cancel the normal part of the applied field, so that the total surface field $\vector E_s$ is tangent to the surface (see Fig. \ref{fig:sketch}a).   Evidently the net depolarization charge is zero, so that the depolarization field at infinity has in general a dipolar form.  The depolarization field is necessarily present whenever there is an applied field that produces current; it is independent of any static charge in or near the object.  Such static charges produce no current and thus create no depolarization field. 
	
\subsection{				Charged body strongly screened with countercharge}
\begin{figure}
\includegraphics[width=.5\textwidth]{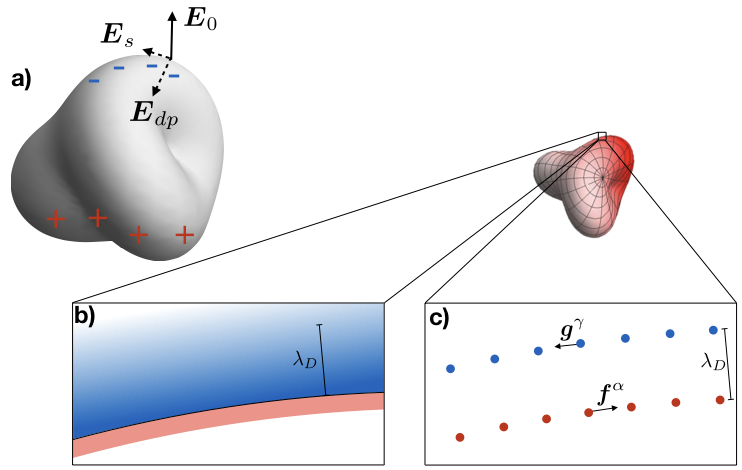}
\caption{\label{fig:sketch} a) Depolarization charge, shown as pluses and minuses, on an insulating body in a conducting medium in the presence of an applied field $\vector E_0$. The charge creates a field $\vector E_{dp}$ which cancels the normal component of $\vector E_0$ at the surface, as no current (and therefore no field) can penetrate the body. The resultant field $\vector E_s$ is the total field at the object's surface. b) Physical picture of electrophoresis. Counterions in the surrounding fluid form a screening cloud (blue) of thickness $\lambda_D$, which act to screen the surface charge (red) of the object. c) Stokeslet picture of electrophoresis. Both the body surface charges (red dots) and screening charges (blue dots) are represented as point sources of drag called stokeslets, and exert respective drag forces $\vector f^\alpha$ and $\vector g^\gamma$ on the fluid. The two layers of stokeslets are separated by a length $\lambda_D$.}
\end{figure}

It is known that when a charged colloidal body is suspended in an electrolytic solution, the mobile counterions in the solution form a screening double layer that acts to cancel the charge on the body, as shown in Fig. \ref{fig:sketch}b. This screening layer has a thickness $\lambda_D$ defined above which depends on factors such as ion concentration and temperature.  It is set by the balance between electrostatic and thermodynamic forces; in the linear-response regime of interest these are negligibly perturbed by the applied field $\vector E_0$. We only consider cases where $\lambda_D$ is much smaller than the object. There is an electric potential $\zeta$ set up in the screening layer, known as the zeta potential, which depends on $\lambda_D$ and the surface charge density $\sigma$:
\begin{equation}
	\zeta = \frac{\sigma\lambda_D}{\epsilon}
\end{equation}
where $\epsilon$ is the dielectric permittivity of the fluid. If we consider a {\it uniformly} charged body of size $R$ in electrophoresis, its motion is independent of the shape of the body in the limit $\lambda_D\muchlessthan R$.  The velocity $\vector V$ is given by $\vector V = \mu \vector E_0$ where the coefficient $\mu$ is called the mobility \cite{Morrison}:
\begin{equation} \label{smol}
	\mu = \frac{\epsilon\zeta}{\eta},
\end{equation}
where $\eta$ is the viscosity of the fluid. A uniformly charged object will move in the direction of the applied field and will not rotate, so the mobility $\mu$ completely characterizes its electrophoretic response.

When a body has non-uniform charge, the zeta potential depends on the position $s$ on the body's surface.  We can no longer describe the motion with a single scalar quantity. Instead, the motion is characterized by translational and rotational mobility tensors $\AA$ and $\TT$, where $\vector V = \AA\vector E_0$ and $\vector \Omega = \TT\vector E_0$. It is useful to normalize $\zeta(s)$ (as well as $\mu(s)$ and $\sigma(s)$) by their rms values, \eg $\sqrt{\expectation{\zeta^2}}$, thus defining \eg $\tilde \zeta(s) \definedas \zeta(s)/\sqrt{\expectation{\zeta^2}}$.  We then can define reduced mobility tensors $\tilde{\AA}$ and $\tilde{\TT}$ such that
\begin{gather}
	\vector V = \sqrt{\expectation{\mu^2}}~\tilde{\AA}~\vector E_0 \label{eq:tildeTransMobility} \\
	\vector \Omega = \frac{\sqrt{\expectation{\mu^2}}}{R}~\tilde{\TT}~\vector E_0 \label{eq:tildeRotMobility}
\end{gather}
where $R$ is the Stokes radius of the body \cite{StokesRadius}. This definition guarantees the mobility tensors will be invariant under rescalings of the object size or charge density.

For a non-uniformly charged sphere, these tensors are given by \cite{AndersonSpheres}
\begin{gather}
	\tilde{\AA} = \expectation{\tilde\zeta} \mathbb{I} - \frac{1}{2}\tilde{\mathbb{Q}} \label{eq:force}\\
	\tilde{\TT} = \frac{9}{4} \crossmat{\tilde{\vector p}}\label{eq:torque}
\end{gather}
where the matrix $\crossmat{\tilde{\vector p}}$ is defined such that $\crossmat{\tilde{\vector p}} \vector E_0 = \tilde{\vector p}\times\vector E_0$, $\mathbb{I}$ is the identity matrix, and $\expectation{\tilde\zeta}, \tilde{\vector p}$, and $\tilde{\mathbb{Q}}$ are the normalized monopole, dipole, and quadrupole moments, respectively, of the zeta potential:
\begin{gather}
	\expectation{\tilde\zeta} = \frac{1}{4\pi}\iint\tilde\zeta(s)\, d\Omega \\
	\tilde{\vector p} = \frac{1}{4\pi}\iint\tilde\zeta(s) \hat{\vector n}(s)\, d\Omega \label{eq:pdot}\\\
	\tilde{\mathbb{Q}} = \frac{1}{4\pi}\iint\tilde\zeta(s) \left(3\hat{\vector n}(s)\hat{\vector n}(s) - \mathbb{I}\right) d\Omega \label{eq:Qtilde}
\end{gather}
where $\hat{\vector n}(s) $ is the unit vector normal to the sphere's surface at $s$.   We note that Eq. \eqref{eq:torque} describes the familiar response of a dipole.  The object rotates transiently until its dipole moment is aligned with the applied field.   However, the  translational motion is more distinctive.  Steady-state translational motion is not in general parallel to $\vector E_0$.
			
\subsection{				Conservation laws}
\label{sec:conservation}
We noted above that electrophoresis has a qualitatively different effect on the surrounding fluid than sedimentation has.  We define everything within the screening layer to be a mechanical system.  This system emits no electric field to the exterior.  Having fixed these charges, we now apply an exterior field such as $\vector E_0$.  Since the system is equivalent to an uncharged body, there can be no net force or torque on it.  If one considers the perturbation of the system charge owing to $\vector E_0$, the resulting polarization effects can exert forces or torques.  But these forces are of order $\vector E_0{}^2$ and are not considered here.  

Though there can be no net force, there can be local forces that add to zero.  Indeed, the translational motion of the body must displace the outer (incompressible) fluid, requiring viscous work.  The needed displacement field is that required to remove the object's volume at the original position and then insert it at its translated position.  This displacement field is a potential flow known as a mass dipole \cite{Diamant:2009sj}.  Since the monopole displacement falls off as inverse distance squared, the mass dipole field, being the gradient of the monopole, falls off as the inverse distance to the third power, in contrast to the long-range inverse-distance dependence of the Oseen flow.  

\section{					Discrete-particle calculation of tensor electrophoresis}
\label{sec:electrophoresis}
\subsection{				Representation of body}
We now discuss how we can apply the stokeslet method described above to electrophoresis.  We first form a stokeslet object made of many stokeslets positioned to conform to the shape of the body, as in Section \ref{sec:stokeslets}.  We represent the given charge density profile over the body by assigning discrete charges to these stokeslets.  We must have enough stokeslets so that the shape and charge of the object are properly resolved, though the stokeslets can be as dilute as we like if they are numerous enough, as discussed in Sec. \ref{sec:stokeslets}.

\subsection{				Representation of screening charge}\label{sec:screeningCharge}
We also need to represent the diffuse charge in the fluid that screens the object. In the small Debye length limit, this charge falls off exponentially, with screening length equal to $\lambda_D$.  We instead represent the screening charge as a shell of charged stokeslets. For simplicity we create one screening stokeslet for each of the charged body stokeslets, displaced from its counterpart at a distance $\lambda_D$ normal to the surface, as illustrated in Fig. \ref{fig:sketch}c.  The static screening charge distributes itself so as to minimize the electrostatic energy.  Accordingly, we determine the charge $q_\gamma$ on each screening stokeslet $\gamma$ by minimizing the electrostatic energy of the system in the presence of the fixed body charges.  This minimization necessarily removes exterior multipole fields.  

\subsection{				Forces on a stokeslet or charge}

The aim of our method is to account for all the forces on the fluid arising from an applied electric field $\vector E_0$.  This includes forces both from the body and from the screening charge.  We first distinguish the forces that produce motion, proportional to $\vector E_0$, from those that do not.

\subsubsection{			Forces in the absence of applied field}

In the absence of applied fields, the system is in equilibrium, with no flow and no drag forces. The strong electric field in the screening layer is balanced by osmotic pressure that maintains the separation $\lambda_D$.  These strong fields are presumed to be indefinitely larger than the small electrophoretic field $\vector E_0$.  Thus the static charge profiles are presumed to be disturbed by $\vector E_0$ to a negligible degree.  Likewise, these charge profiles are presumed to be disturbed negligibly by the flow resulting from $\vector E_0$.  

\subsubsection{			Forces linear in applied field}

When the field $\vector E_0$ is applied, other forces arise in proportion to it.  As we have seen, the depolarization field arises so that the net electric field is tangent to the surface.  The velocity field and drag forces are also proportional to $\vector E_0$.

Part of the external field acts on the body charges. We need not account for these individually; since the body is rigid, the only forces that affect its motion are the net force and torque on the body.  These we determine below.

Each screening charge $q_\gamma$ transmits to the fluid a drag force $\vector g^\gamma$ equal to the net non-drag force acting on it, shown in Fig. \ref{fig:sketch}c.  This net force includes the electrostatic force $q_\gamma \vector E(\vector r^\gamma)$.  It also includes an equilibrium force that maintains the screening layer at fixed separation from the body, forcing a stokeslet at $\vector r^\gamma$ to have a velocity normal to the surface equal to the body's: $\hat{\vector n} \cdot \vector U^\gamma = \hat{\vector n} \cdot (\vector V + \vector \Omega \times \vector r^\gamma)$.  The corresponding constraint force also acts only normal to the surface.  As with the body stokeslets, the full drag force on the fluid is proportional to the velocity of the adjacent fluid $\vector v(\vector r^\gamma)$ relative to that of the stokeslet itself $\vector U^\gamma$:
\begin{equation}
\vector g^\gamma = -\Gamma \left [\vector v(\vector r^\gamma) - \vector U^\gamma \right ]
\end{equation}
But unlike the body stokeslets, only part of the velocity $\vector U^\gamma$ is constrained: \viz the normal component $\vector U^\gamma \cdot \hat{\vector n}$.  The tangential part of $\vector g^\gamma$, denoted $\mathcal T \vector g^\gamma$, is given by the electric force. Thus, 
\begin{eqnarray}
\mathcal T(\vector r^\gamma) ~\vector g^\gamma =& \mathcal T(\vector r^\gamma)  ~q_\gamma \vector E(\vector r^\gamma) \\
\hat{\vector n}(\vector r^\gamma)\cdot \vector g^\gamma  =& -\Gamma \left [\hat{\vector n}(\vector r^\gamma)
\cdot \vector v(\vector r^\gamma) -  \hat{\vector n}(\vector r^\gamma)\cdot\vector U(\vector r^\gamma) \right ] \nonumber \\
 =&~~  -\Gamma \left [\hat{\vector n}(\vector r^\gamma)\cdot \vector v(\vector r^\gamma)
 - \hat{\vector n}(\vector r^\gamma)\cdot (\vector V + \vector \Omega \times \vector r^\gamma) \right ]\label{eq:normalpart}\nonumber \\
 \end{eqnarray}

These equations for $\vector g^\gamma$ can be simplified further.  We have noted that the role of the normal force $\hat{\vector n} \cdot \vector g^\gamma$ is to satisfy a constraint: the screening charge cloud must remain at a fixed distance from the body.  The force on the stokeslet is supplied by the osmotic force on the nearby body.  If the component of another force (\eg the $\vector E$ field) would violate the constraint, the osmotic force cancels this component.  There is an equal and opposite force on the body nearby.  This pair of constraint forces is an internal action-reaction pair and can have no impact on the total force or torque.  This constraint force may thus be ignored just as we ignored the constraint forces that maintain the body's shape.  We may thus eliminate the normal component of $\vector g^\gamma$ in Eq. \eqref{eq:normalpart}.

\subsection{Self-consistent drag forces }
\label{sec:selfconsistent}
As with sedimentation, each stokeslet force $\vector f^\alpha$ must be consistent with its motion relative to the fluid: $\vector f^\alpha = -\Gamma~(\vector v(\vector r^\alpha) - \vector U^\alpha )$.  The fluid velocity $\vector v$ includes contributions from the body stokeslets as in Eq. \eqref{eq:fConsistency}.  It also includes a contribution from the screening charge stokeslets, \viz $\sum_\gamma \GG(\vector r - \vector r^\gamma) \vector g^\gamma$. Having dropped the constraint force $\hat{\vector n} \cdot \vector g^\gamma$ as justified above, the self-consistency equations for the body forces $\vector f^\alpha$ and screening forces $\vector g^\gamma$ become 
\begin{eqnarray}
\vector f^\alpha =& -\Gamma \bigg (
\sum_{\beta \notequal \alpha} \GG(\vector r^\alpha - \vector r^{\beta}) ~ \vector f^\beta  \nonumber\\*
&+ \sum_\gamma \GG(\vector r^\alpha - \vector r^\gamma) \vector g^\gamma \nonumber \\*
&- \vector V - \vector \Omega \times \vector r^\alpha \bigg) \nonumber\\*
\vector g^\gamma  =& \mathcal T(\vector r^\gamma)~q_\gamma \vector E(\vector r^\gamma) \nonumber\\*
 \label{eq:qConsistency}
\end{eqnarray}

As with Eq. \eqref{eq:fConsistency}, these equations determine the drag forces $\vector f$ and $\vector g$ for any assumed $\vector E_0$ and velocities $\vector V$ and $\vector \Omega$.  However, it doesn't determine $\vector V$ and $\vector \Omega$ themselves.  It also makes no reference to the electric forces on the body.  Nevertheless, we may determine $\vector V$ and $\vector \Omega$ by calculating the net force on the fluid, \ie the sum of the Stokes drag forces $\vector f$ and $\vector g$.  As noted in Section \ref{sec:conservation}, this sum must vanish.  This requirement imposes a linear constraint on $\vector V$ and $\vector \Omega$.  The drag forces must also exert no torque on the fluid.  This requirement imposes a second constraint on $\vector V$ and $\vector \Omega$.  Taken together, these constraints determine the $\vector V$ and $\vector \Omega$ for the given imposed $\vector E_0$, and thence the mobility matrices $\AA$ and $\TT$.  
			
\section{					Numerical implementation }
\label{sec:spheres}
We implemented the procedure above for a sphere with various distributions of charge in order to assess its validity and feasibility.  In this section we provide the specifics of our calculation not covered above.  Our aim was to do all the calculations in a way that would be equally feasible for an asymmetric shape.  We describe how the stokeslets are positioned,  how the countercharge magnitudes are determined,  how the net electric field $\vector E(\vector r)$ is determined, and how the self-consistency equations are solved.

\subsection{				Distributing body stokeslets, countercharge stokeslets}
We first need to specify how we arrange the stokeslets to represent our object and its screening layer.  We chose two levels of refinement, one with 499 body stokeslets and one with 1999.  In both cases the stokeslets were distributed on a sphere according to the Spiral Points algorithm of Rakhmanov, Evguenii, Saff and Zhou \cite{Rakhmanov:1994fk, Kuijlaars:1998uq}.  We determined that the 1999-stokeslet sphere behaved like a hydrodynamically solid sphere to good accuracy as shown in the next section.   We then assigned a charge magnitude to each stokeslet so that our object had the desired surface charge density profile.  Once we had distributed our body stokeslets, we placed a screening charge stokeslet a distance $\lambda_D$ (the Debye length) from each body charge normal to the object's surface, giving us a shell of screening stokeslets.  The magnitudes of these screening charges are determined next.

\subsection{				Screening charge distribution}
As mentioned before, the screening charge acts to cancel the equilibrium electric field of the object.  Therefore, we vary the screening charge magnitudes $q_\gamma$ while keeping their positions fixed so as to minimize their electrostatic energy in the presence of the body charges, effectively treating the screening layer as a grounded conducting shell.  Thus we may determine the charge state by requiring that the total electrostatic  potential at each charge $\gamma$ vanish, \ie for each $\gamma$:
\begin{equation}
	\sum_{\beta\neq\gamma} \frac{q_{\beta}}{\left|\vector r^{\gamma}-\vector r^{\beta}\right|} + Cq_{\gamma} + \Phi\left(\vector r^{\gamma}\right) = 0,
\label{eq:phiofq}
\end{equation}
where $q_{\beta}$ are the screening charge magnitudes, $\vector r^{\beta}$ are their positions (held fixed), and $\Phi\left(\vector r\right)$ is the electrostatic potential at $\vector r$ due to the body charges.  We include a nonzero ``self-potential" contribution $C q_\gamma$ to the potential at $\vector r^\gamma$ due to the charge $q_\gamma$ itself. This self potential is present for any smooth charge distribution approximated by discrete charges; it becomes negligible compared to the first term in the limit of many small charges, but including it gives greater accuracy for a finite number of charges.  We choose $C$ by a simple empirical criterion \footnote{We choose $C$ by solving Eq. \eqref{eq:phiofq} for a uniformly charged sphere. Any choice of $C$ gives a set of screening charges $q_\gamma$.  We choose $C$ such that the total screening charge is equal and opposite to the total body charge.  $1/C$ is expected to be the capacitance of a small disk whose diameter is roughly the distance between stokeslets.  Our empirical choice of $C$ is in good agreement with this expectation.}.   

\subsection{				Depolarization field}\label{sec:depolarization}

The total external field $\vector E(\vector r)$ (applied plus depolarization) can be found analytically for highly symmetric cases. For an insulating sphere of radius $R$ \cite{AndersonEllipsoids}, this field is given by
\begin{equation} \label{eq:depfield}
	\vector E\left(\vector r \right) = \left(1+\frac{R^3}{2r^3}\right)\vector E_0 - \frac{3R^3}{2r^3}\left(\vector E_0\cdot\hat{\vector r}\right)\hat{\vector r}.
\end{equation}
This is the external field that all charged stokeslets feel.

Finding the depolarization field for arbitrary shapes becomes difficult analytically, as one must solve the Laplace equation with a Neumann boundary condition at the object's surface. We outline an alternative computational approach in Appendix \ref{sec:numEfield} that fits naturally with the stokeslet hydrodynamics described above. However, our results in Section \ref{sec:results} simply use Eq. \eqref{eq:depfield} for the depolarization field and do not test this method.

\subsection{				Velocity and angular velocity}

Now the quantities for Eq. \eqref{eq:qConsistency} are known, so we can determine the drag forces \vector f for any given applied field $\vector E_0$ and velocities $\vector V$ and $\vector \Omega$. However, as described in Section \ref{sec:selfconsistent}, Eq. \eqref{eq:qConsistency} does not explicitly determine the velocities. We therefore find the drag forces in terms of $\vector V$ and $\vector \Omega$. We then impose the condition of zero net force by expressing $\sum_\alpha \vector f^\alpha + \sum_\gamma \vector g^\gamma$ using the right side of Eq. \eqref{eq:qConsistency} and setting this total equal to zero, thus giving three linear constraints on $\vector V$ and $\vector \Omega$.  We also impose the condition of zero net torque by expressing $\sum_\alpha \vector f^\alpha \times \vector r^\alpha + \sum_\gamma \vector g^\gamma \times \vector r^\gamma$, again using the right side of Eq. \eqref{eq:qConsistency}, and setting this total equal to zero, thus giving three further constraints on $\vector V$ and $\vector \Omega$. These six constraints together determine \vector V and \vector \Omega, as desired, and thence the stokeslet forces. Now the fluid velocity $\vector v(r)$ is simply the sum of the stokes velocities for each stokeslet, \ie
\begin{equation}
\vector v(\vector r) = \sum_\alpha \GG(\vector r - \vector r^\alpha) \vector f^\alpha 
+ \sum_\gamma \GG(\vector r - \vector r^\gamma) \vector g^\gamma 
\end{equation}

\section{					Results}\label{sec:results}

\begin{figure}
\includegraphics[width=.5\textwidth]{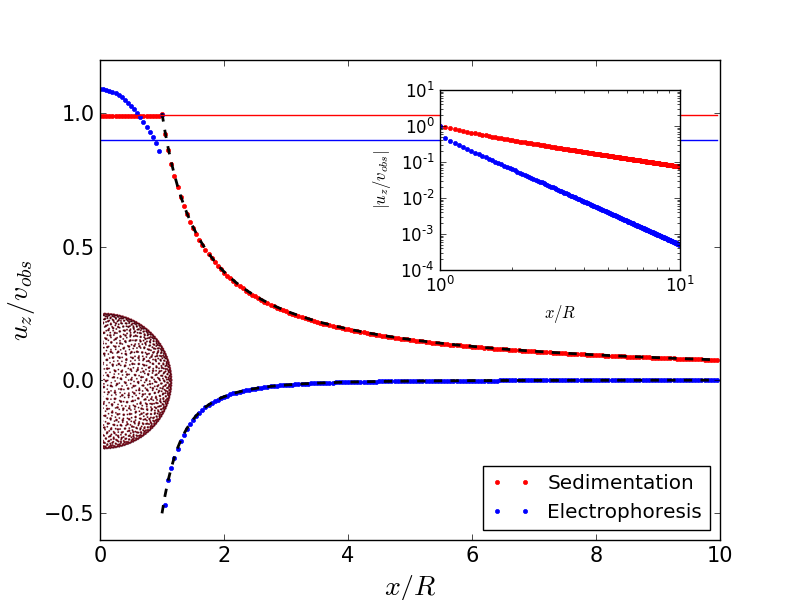}
\caption{\label{fig:sedelph}Comparison of the velocity field from a sedimenting 1999-stokeslet sphere (red dots) and a uniformly charged electrophoretic sphere with $\lambda_D = 3\%$ of the sphere radius (blue dots), using the methods of Section \ref{sec:spheres} (sphere and stokeslets are shown to scale).  Vertical velocity of the fluid $u_z$, normalized by the calculated body speed $v_{obs}$, is plotted versus horizontal distance $x$ from the center. For the sedimenting case, the fluid within the sphere is moving with the object to within a $1 \%$ tolerance (horizontal red line) for $x < 1$.  For $x>1$ the velocity falls off as $1/r$ as shown in the log-log inset plot, and agrees well with theoretical predictions \cite{acheson} (dashed line).   The speed of the sphere implies a Stokes drag coefficient about $1 \%$ smaller than the theoretical value for a sphere (red line).  For electrophoresis, the fluid near the center moves somewhat faster than the sphere. External flow varies as $k /r^3$ as expected for mass dipole flow.  The expected value of $k$ is 1/2 \cite{AndersonSpheres}, indicated by the dashed line. The external velocity field matches that expected for a solid sphere moving at the observed velocity to within $1 \%$.  This observed velocity in turn agrees with the predicted velocity (blue line) to $11 \%$, \cf Table \ref{nonuniformmobilitytable}.}
\end{figure}

We now demonstrate that our stokeslet method reproduces all of the characteristics of electrophoresis described above. As mentioned before, we find that a sphere made up of a single layer of 1999 stokeslets moves at the correct speed when a constant force is applied (as in sedimentation) as shown in Table \ref{sedmobilitytable}, and that the flow in the interior moves with the object, as shown in Fig. \ref{fig:sedelph}. We also find that a uniformly charged stokeslet sphere with a surrounding stokeslet screening layer in electrophoresis (as described in the previous section) moves in the direction of the applied field with approximately the proper electrophoretic mobility, as shown in Table \ref{elphmobilitytable}. Furthermore, we see that the sedimentation flow falls off as $1/r$, like a force monopole, while the electrophoresis flow falls off as $1/r^3$, like a mass dipole, shown in Fig. \ref{fig:sedelph}. We also compute the mobility tensors for various non-uniformly charged stokeslet spheres, and see that they agree with the known values to within less than 4\% (Table \ref{nonuniformmobilitytable}). Discrepancies from theoretical predictions decrease as numerical approximations improve, as we now discuss.

\begin{table}[h]
\begin{tabular}{| c | c | c | c | c |}
    \hline
    $N$ & $a$ & Predicted Mobility & Measured Mobility & Error \\ \hline \hline
    499 & 0.03 & $1/(6\pi) = 0.05305$ & 0.05395 & 1.7\% \\ \hline
    1999 & 0.015 & 0.05305 & 0.05351 & 0.9\% \\ \hline
\end{tabular}
\caption{\label{sedmobilitytable} Sedimentation mobilities of spheres. Predicted mobility is given by $1/(6\pi\eta R)$. As the number of stokeslets $N$ is increased, the error in the mobility decreases, indicating that using more stokeslets gives more accurate results.}
\end{table}

\begin{table*}
\begin{tabular}{| c | c | c | c | c | c | c | c | c |}
    \hline
    $N$ & $a$ & $\lambda_D$ & $\phi_s$ & $\xi_e$ & Predicted Mobility \cite{OhshimaMobility} & Measured Mobility & Error \\ \hline \hline
    499 & 0.03 & 0.06 & 0.45 & 0.07 & $4.309\cdot10^{-3}$ & $4.736\cdot10^{-3}$ & 10\% \\ \hline
    1999 & 0.015 & 0.06 & 0.45 & 0.03 & $4.309\cdot10^{-3}$ & $4.455\cdot10^{-3}$ & 3.4\% \\ \hline
\end{tabular}
\caption{\label{elphmobilitytable} Electrophoretic mobilities of uniformly charged spheres, where total surface charge is 1. The nominal penetration depth $\xi_e$ is defined in Appendix \ref{sec:screening}.  In the second row, the reduced size of the stokeslets is compensated by an increased number of stokeslets, so that the penetration depth $\xi_e$ decreases.  The result is a reduced error in the mobility.}
\end{table*}

\begin{table}[h]
\begin{tabular}{| c | c | c | c | c |}
    \hline
    \shortstack{Charge \\ distribution} & \raisebox{.5\height}{$\lambda_D$} & \shortstack{Predicted \\ Eqs. \eqref{eq:force} - \eqref{eq:Qtilde}} & \raisebox{.5\height}{Measured} & \raisebox{.5\height}{Error} \\ \hline \hline
    \multirow{2}{*}{Uniform} & 0.03 & \raisebox{-.25\height}{$\tilde \AA =  1$} & 1.053 & 5.3\% \\ \cline{2-5}
	& 0.06 & \raisebox{-.25\height}{$\tilde \AA =  1$} & 0.933 & 6.7\% \\ \hline
    \raisebox{-.5\height}{\shortstack{Capped \\ \includegraphics[width=.15\textwidth]{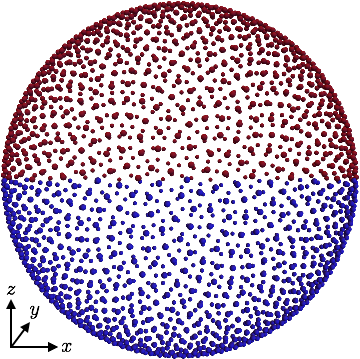}}} & 0.03 & $\tilde \TT_{yx}$ = 1.125 & 1.1621 & 3.3\% \\ \hline
    \raisebox{-.5\height}{\shortstack{Dipolar \\ \includegraphics[width=.15\textwidth]{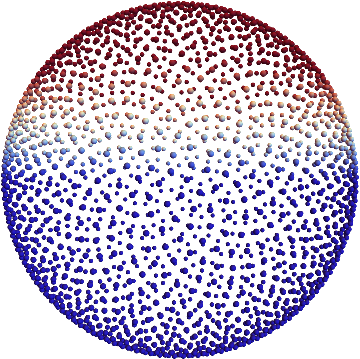}}} & 0.03 & $\tilde \TT_{yx}$ = 1.299 & 1.3419 & 3.3\% \\ \hline
    \raisebox{-.5\height}{\shortstack{Striped \\ \includegraphics[width=.15\textwidth]{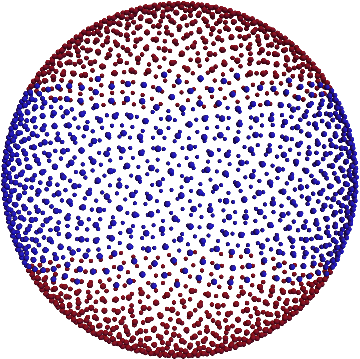}}} & 0.03 & \raisebox{-.25\height}{\shortstack{$\tilde \AA_{xx}$ = 0.1875 \\ $\tilde \AA_{zz}$ = -0.375}} & \raisebox{-.25\height}{\shortstack{0.189 \\ -0.3697}} & \raisebox{-.25\height}{\shortstack{0.8\% \\ 1.4\%}} \\ \hline
    \raisebox{-.5\height}{\shortstack{Quadrupolar \\ \includegraphics[width=.15\textwidth]{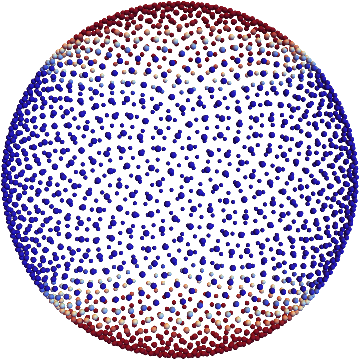}}} & 0.03 & \raisebox{-.25\height}{\shortstack{$\tilde \AA_{xx}$ = 0.2236 \\ $\tilde \AA_{zz}$ = -0.4472}} & \raisebox{-.25\height}{\shortstack{0.2221 \\ -0.4444}} & \raisebox{-.25\height}{\shortstack{0.7\% \\ 0.6\%}} \\ \hline
\end{tabular}
\caption{\label{nonuniformmobilitytable} Reduced electrophoretic mobilities (Eqs. \eqref{eq:force}, \eqref{eq:torque}) for spheres with different charge distributions. Predicted mobilities are for $\lambda_D \goesto 0$ in contrast to Table \ref{elphmobilitytable}.  Measured mobilities were for $N = 1999$ and $a = 0.015$. The mobilities reported for the capped and dipolar spheres are $\tilde \TT_{yx}$ (so these spheres may rotate), while the mobilities reported for striped and quadrupolar spheres are $\tilde \AA_{xx}$ and $\tilde \AA_{zz}$ (so these spheres may translate obliquely to the applied field). Unreported elements of $\tilde \AA$ and $\tilde \TT$ are predicted to be 0 and measured to be much smaller than those shown.}
\end{table}

\begin{figure}
\includegraphics[width=.5\textwidth]{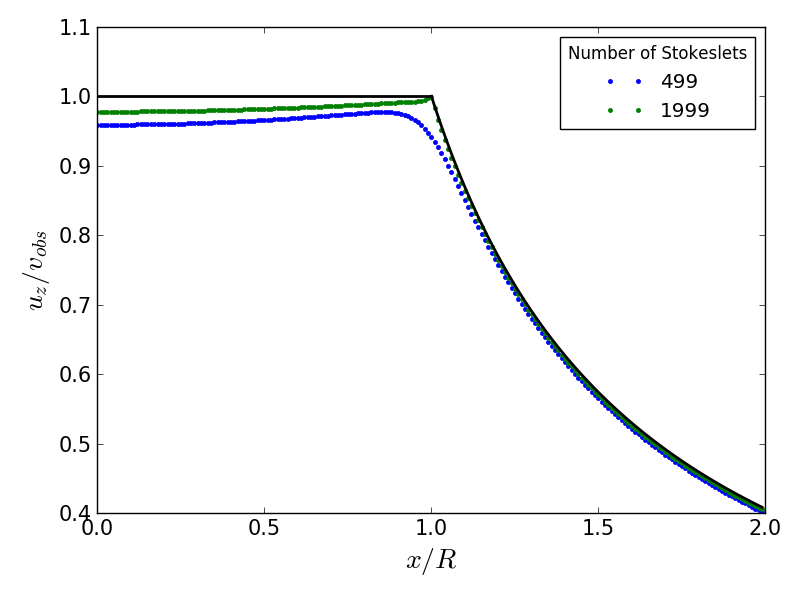}
\caption{\label{fig:sedstokesletnumdep}Effect of number of stokeslets on sedimentation flow, plotted as in Fig. \ref{fig:sedelph}. Inset indicates the number of stokeslets for each data set. When more stokeslets are used to represent a rigid body, better agreement with theory is obtained, as shown in Table \ref{sedmobilitytable}. Furthermore, the transition from the interior to the exterior is sharper for 1999 stokeslets than for 499, as the sphere blocks the flow better with more stokeslets. The solid black line corresponds to the flow profile of a perfect hard sphere.}
\end{figure}

\begin{figure}
\includegraphics[width=.5\textwidth]{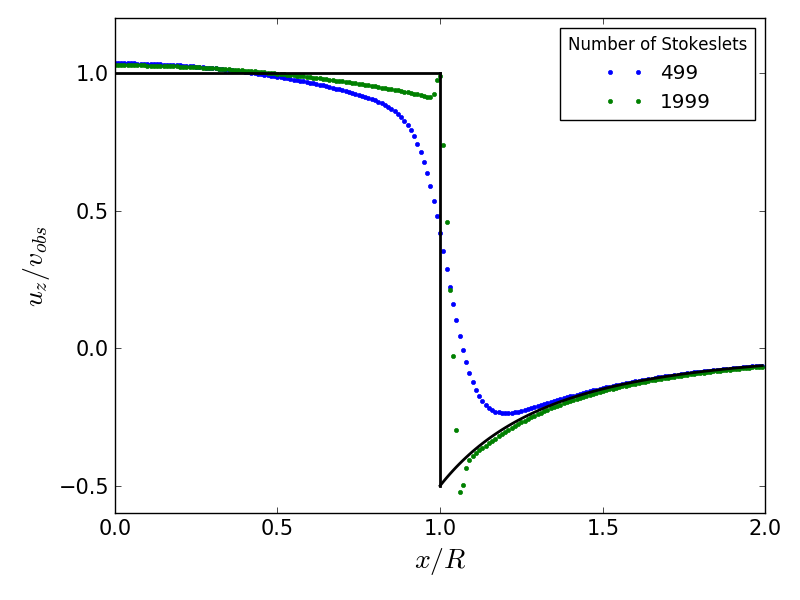}
\caption{\label{fig:stokesletnumdep}Effect of number of stokeslets on the electrophoretic flow, plotted as in Fig. \ref{fig:sedstokesletnumdep} ($\lambda_D = 6\%$ of $R$ in both cases).  When more stokeslets are used to represent a rigid body, better agreement with theory is obtained, as shown in Table \ref{elphmobilitytable}. Furthermore, the transition from the interior to the exterior is sharper for 1999 stokeslets than for 499, as the sphere blocks the flow better with more stokeslets. The solid black line corresponds to the flow profile of a perfect hard sphere with zero electrostatic screening length.}
\end{figure}

\begin{figure}
\includegraphics[width=.5\textwidth]{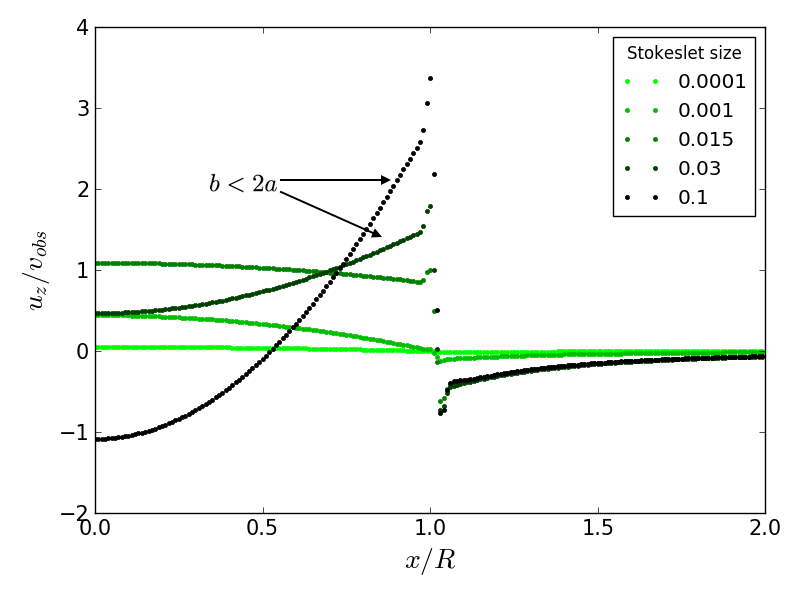}
\caption{\label{fig:stokesletsizedep}Effect of stokeslet size $a$ on the electrophoretic flow, plotted as in Fig. \ref{fig:sedstokesletnumdep}, for $N = 1999$ and $\lambda_D = 0.03$. For very small $a$, the fluid inside the sphere moves much slower than the sphere itself. As $a$ increases, the interior flow matches better with the sphere velocity, so the sphere becomes more hydrodynamically opaque. However, when $a$ is greater than half the nearest-neighbor distance between stokeslets $b$ or $\lambda_D$, stokeslets begin to overlap, and the flow becomes unphysical.}
\end{figure}

\begin{figure}
\includegraphics[width=.5\textwidth]{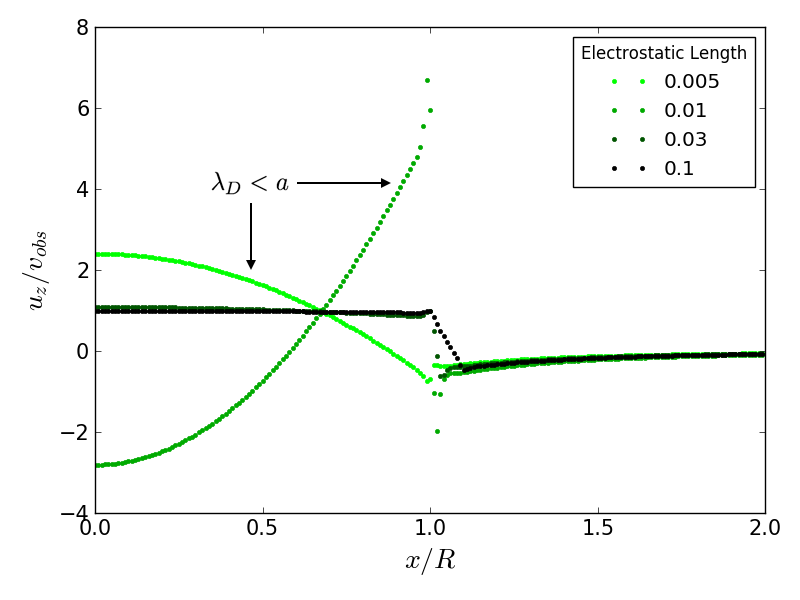}
\caption{\label{fig:lambdadep}Effect of electrostatic screening length $\lambda_D$ on the electrophoretic flow, plotted as in Fig. \ref{fig:sedstokesletnumdep}, for $N = 1999$ and $a = 0.015$. When $a < \lambda_D \ll R$, the interior flow matches well with the sphere velocity. However, when $\lambda_D < a$, the object and screening charge stokeslet layers begin to overlap, and the flow becomes unphysical.}
\end{figure}

\begin{figure}
\includegraphics[width=.5\textwidth]{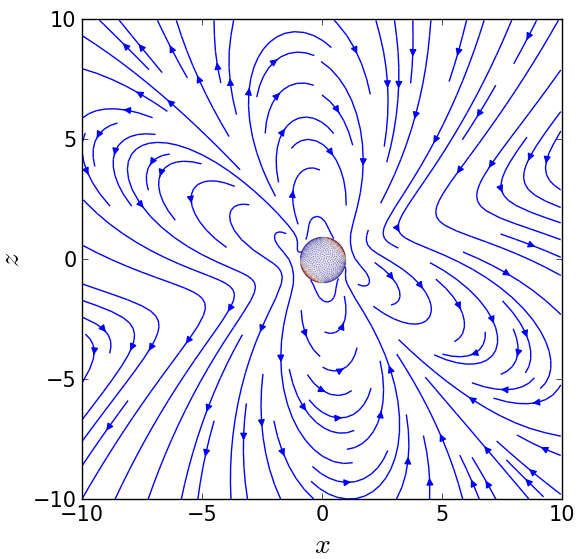}
\caption{\label{fig:quadstream} Streamlines for the quadrupolar sphere of Table \ref{nonuniformmobilitytable} tilted 45 degrees to the right under an applied field in the $z$-direction. The sphere and its surface charge are shown at scale.  The streamlines indicate only the direction of the flow, not its magnitude.}
\end{figure}

\subsection{				Error versus number of stokeslets}
The theory treats a solid sphere with uniform charge in the limit of small Debye screening length. Our numerical method approaches this limit when the number of stokeslets is large and the screening charge is close to the sphere. We see that the flow inside the sphere is strongly suppressed, both in sedimentation and electrophoresis, again shown in Fig. \ref{fig:sedelph}. We also find that as we increase the number of stokeslets from 499 to 1999, the error in the calculated velocity of the object decreases, both in sedimentation (Table \ref{sedmobilitytable}) and electrophoresis (Table \ref{elphmobilitytable}). In addition, we get better hydrodynamic screening with more stokeslets, shown in Figures \ref{fig:sedstokesletnumdep} and \ref{fig:stokesletnumdep}. When we increase the size of our stokeslets, as in Fig. \ref{fig:stokesletsizedep}, we see that our stokeslet sphere becomes more opaque, and the flow inside the sphere matches its speed. We also can change the Debye length $\lambda_D$, and see that the flow inside the sphere matches its velocity when $\lambda_D/R$ is small (Fig. \ref{fig:lambdadep}). However, in cases where the stokeslet size is larger than either the Debye length or average stokeslet spacing $b$, there are stokeslets which are overlapping, which is geometrically unphysical.
			
\subsection{				Scaling of computer effort vs number of stokeslets $N$}
The calculations above require computing a matrix $\mathbb{M}$ of dimensions $3N \times 3N$ and then solving the linear equation \eqref{eq:qConsistency}.  For very large $N$, the latter step must dominate the computer effort; it is expected to grow faster than quadratically in $N$ \cite{golub2013matrix}.  However, for the $N$ values of our study, the computer time was always dominated by the time to compute the matrix $\mathbb{M}$, quadratic in $N$.  Considering Eq. \eqref{eq:qConsistency} alone, the required time was not significantly improved when the real $\mathbb{M}$ was replaced by a random, symmetric $\mathbb{M}$. Specifically, for $N = 1999$ the random matrix calculation was 1.5 times faster than for the stokeslet system (on a a 2.6 GHz quad-core MacBook Pro with 16 GB of RAM, using Python 2.7.6). Decreasing the number of stokeslets to 499 resulted in a factor 18 decrease in computer time. The computer time needed for the other cases calculated was consistent with these times.
\subsection{				Observed flow around multipolar charges}
As described before, we can compute the flow field around our sphere for any charge distribution. We find that the flow around a sphere with a quadrupolar surface charge density falls off as $1/r^3$, consistent with a force quadrupole. Furthermore, we compute the streamlines around a quadrupolar sphere, shown in Fig. \ref{fig:quadstream}.
\subsection{				Mobilities}
Finally, we compare the known electrophoretic mobilities for several charge distributions against those obtained by our stokeslet method. The objects studied are pictured in Table \ref{nonuniformmobilitytable}.  We considered only charge distributions with monopolar, dipolar, or quadrupolar charge distributions, since \cite{AndersonSpheres} higher multipoles give vanishing mobility.  We compared pure dipole and quadrupole charge distributions to step-function distributions with the same symmetry in order to gauge the effect of abrupt changes in charge density on our numerical accuracy.  Table \ref{nonuniformmobilitytable} reports our results in terms of the normalized tensors $\tilde \AA$ and $\tilde \TT$.  We chose an electrostatic screening length $\lambda_D$ to be .03 $R$ and a stokeslet radius $a$ of .015 corresponding to a realistic internal flow as shown in Figure \ref{fig:stokesletsizedep}. In all the cases studied, the mobilities proved to be well within less than $4 \%$ of the expected values.  The lowest accuracy was found for the dipolar charge distributions.  The pure dipole and its step-function counterparts showed virtually the same accuracy; the same was true for the quadrupoles.  The quadrupoles are characterized by two mobilities, $\tilde \AA_{xx}$ and $\tilde \AA_{zz}$.
			
\section{					Discussion}\label{sec:discussion}
\subsection{Limitations}\label{sec:limitations}
The foregoing section shows that the stokeslet method can reproduce the known behavior of spheres with monopolar, dipolar or quadrupolar charge distributions to a few percent accuracy using a conceptually simple, numerically practical calculation.  This study also reveals limitations of the stokeslet method.  Being based on discrete charges, the method is only useful insofar as the true flow around the object can be represented by means of stokeslets.  This means that the stokeslets must be dilute, yet numerous.  They must be dilute because they only represent the flow accurately at distances substantially greater than the stokeslet radius $a$. Thus the distance $b$ between neighboring stokeslets must be much greater than $a$.  They must be numerous because with too few stokeslets, the fluid moves through the ensemble of stokeslets, not around them. Such stokeslets would not represent the solid, hydrodynamically opaque objects we wish to describe.  

We explained in Sec. \ref{sec:stokeslets} that the desired hydrodynamic opacity can be achieved with a dilute set of stokeslets.  We noted that the external flow past a set of fixed stokeslets at a given concentration dies out exponentially with depth into the body with a decay length $\xi$ given in Appendix \ref{sec:screening}.    It remains to judge whether our stokeslet bodies are opaque enough to predict electrophoretic properties.  In electrophoresis, the flow occurs primarily between the body and the screening charge layer, at a distance of the order of the electrostatic screening length $\lambda_D$ from the surface.  To represent flow accurately at this scale, the penetration depth $\xi$ must evidently be sufficiently smaller than $\lambda_D$.  

To assess how well this criterion is met, we must estimate $\xi$ for our stokeslet spheres.  Empirically, we may make this estimate by examining the flow around the sedimenting sphere of Fig. \ref{fig:sedstokesletnumdep}.  For the coarser sphere with $N=499$ the velocity crosses over from a constant value near $r=0$ to a decaying function for $r > R$ over a transition region of width of about $R/10$.  This suggests a penetration depth of order $R/10$.  This length evidently decreases as $N$ increases, as the figure shows.  This length agrees well with the analytical estimate $\xi_e$ found in Appendix \ref{sec:screening}.  The area fraction $\phi_s$ for this case is 0.45;  the corresponding  nominal penetration depth $\xi_e\definedas a/\phi_s \aboutequal 0.07$.  The other case shown in Figure \ref{fig:sedstokesletnumdep} has $\xi_e \aboutequal 0.03$.  Any transition region  for this curve is too narrow to see.

The agreement with predictions improves when $\xi_e$ decreases, even when the area fraction $\phi_s$ remains constant, as shown in Table \ref{elphmobilitytable}.

Conversely we showed that when the stokeslets are larger than the screening length or if they overlap, the resulting flow patterns show characteristic signs of being unphysical. 

An important simplification in our scheme was to treat the screening charge layer as a shell spaced one screening length $\lambda_D$ from the body.  The mobility depends only on the flow at distances much larger than $\lambda_D$.  Thus the form of the screening charge decay over distances of $\Order(\lambda_D)$ is not expected to be important.  This assumption is consistent with the boundary-layer treatment of Anderson \cite{anderson1989colloid} \footnote{The continuum boundary-layer picture becomes unreliable when the surface charge density is too strong. When the charge on the body produces an electrostatic energy on a screening charge stronger than about $k_B T$, then the continuum assumptions of our treatment become unreliable.  The discrete screening charges can develop liquidlike correlations and strong atomic-scale correlations with the body charges \cite{Markovich:2014eq}.  This additional structure can inhibit motion of the screening charge and thus add additional forces on the screening charges not considered here.}.

The calculations reported in Sec. \ref{sec:results} support this assumption at the few-percent level.  We did not seek more precise confirmation for two reasons.  The first reason involves the intended purpose of this method.  The purpose is to explore the effect of shape and charge distribution on the motion of actual colloidal objects.  For this purpose it suffices to have a reliable, semiquantitative knowledge of the electrophoretic tensor; a few-percent accuracy is sufficient  for this level of understanding.  

The second reason for our limited accuracy is the numerical properties of the method.  Though the method is conceptually simple and general, its convergence properties are not favorable for precision work.  As noted in Section \ref{sec:results}, determining the tensor mobilities requires solving a $3N \times 3N$ matrix equation, where $N$ is the number of stokeslets.  The matrix consists of the long-range Oseen couplings between stokeslets $\alpha$ and $\beta$; it is not sparse.  Thus, computing the matrix and solving the resulting system involve at least $\Order((3N)^2)$ operations.  These calculations largely implement the cancellation of the Oseen flow around the object, thus it may well be necessary to increase the machine precision of the matrix elements if great accuracy is required.

\subsection{Motion of countercharge}\label{sec:screeningChargeMotion}
In real electrophoresis the screening charge is constantly moving.  It diffuses in and out of the screening layer even with no applied field. During electrophoresis it also advects around the colloidal body, spending only a limited time near any given body.  Thus our method appears paradoxical:  we neglect all of this motion and consider only the initial motion of the screening charge from an assumed equilibrium state.  

This neglect is harmless for two reasons.  First, we aim only to determine the motion to lowest (linear) order in the applied field \footnote{In many cases the applied electric field is too strong to justify this linear-response assumption. For example the applied field can distort the screening charge layer from its equilibrium shape. Our method is not applicable as it stands for such cases.}.  Thus the charge distribution is arbitrarily close to equilibrium.  Second, the resulting flow is quasistatic creeping flow, in which the charge distribution at a given moment is sufficient to determine velocities at that moment.  Our prescription determines these velocities given the equilibrium charge distribution.  The unperturbed thermodynamics guarantees that the system retains this equilibrium charge distribution to arbitrary accuracy.

\subsection{Generalization to nonspherical shapes}\label{sec:nonspherical}
Now that the method has been shown to be accurate for the known case of spheres, the way is cleared to generalize it to its intended use to treat asymmetric objects.  This generalization raises issues that did not arise for spheres.  For concreteness we consider the case of a triaxial ellipsoid.  Here one must first consider how to place the stokeslets.  One obvious method is to place them on a sphere and then stretch the three Cartesian axes to form the desired ellipsoid.  However, this method leads to an uneven distribution of stokeslets over the surface, thus giving very little penetration in some regions while giving too much penetration in others.  

A second issue is determining the screening charge distribution.  We addressed this issue in Section \ref{sec:screeningCharge}.  
As it happens, once the positions of the screening charges have been determined, their magnitudes can be found exactly as was done for spheres: one simply minimizes the electrostatic energy.

Finally one must determine the total electric field $\vector E(\vector r)$ around the body. As mentioned in Section \ref{sec:depolarization}, this is not easily feasible for arbitrary shapes. However, one can use a numerical method similar to what we use to determine the screening charge magnitudes, which we describe in Appendix \ref{sec:numEfield}.

							
\section{Conclusion}

We anticipate a wealth of new phenomena when we apply the method above to asymmetric objects.  The qualitative tensorial behavior is known from our prior study of sedimentation \cite{Krapf,Moths-Witten1,Moths-Witten2}.  However, electrophoresis is much more experimentally convenient.  Further, the tensorial response is likely more rich and informative than in sedimentation.  For example, the charge distribution on an object is amenable to control by convenient environmental factors such as pH.  In addition, electrophoresis is likely easier to interpret than sedimentation response.  This is because the flow around an object in electrophoretic motion falls off rapidly with distance, as noted above.  Thus effects of hydrodynamic interaction are much reduced in the electrophoretic case.  We have begun to explore cases of asymmetric objects that show the potential for the striking behaviors anticipated in Refs. \cite{Krapf,AjdariLong}

\section*{Acknowledgements} The authors are grateful to Prof. Haim Diamant and Dr. Tomer Goldfriend for many clarifying discussions throughout the course of this work.  In addition Brian Moths, Menachem Stern, Daniel Hexner, and Zhiyue Lu provided valuable suggestions on early drafts.  This work was supported in part by a grant from the US-Israel Binational Science Foundation and in part by the University of Chicago Materials Research Science and Engineering Center, which is funded by the National Science Foundation under award number DMR-1420709.

\appendix
\section{Representing solid bodies as stokeslet objects}\label{sec:screening}

Here we clarify the criterion for accurately representing a solid body in a fluid as a layer of many small stokeslet spheres distributed over its surface.  We wish to show that the stokeslets may approximate the solid object arbitrarily well while covering an arbitrarily small fraction of the surface.  We consider an object of nominal size $R$ with $N$ stokeslets of radius $a$ over its surface.  Thus the area per stokeslet is of order $R^2/N$, and the area fraction $\phi_s$ is of order $(a/R)^2 N$.   

These stokeslets interact with a moving fluid via their Oseen tensors, Eq. \eqref{eq:Oseen}, proportional to $a$ and falling off as $1/r$ with distance.   As with the Coulomb potential, the cumulative effect of these Oseen flows grows with system size.  Still, if $a$ is small enough, the perturbation due to these Oseen responses is negligible.  The velocity experienced by a given stokeslet is then the external velocity $v_0$.  The combined effect of the resulting Oseen velocities $\Delta v$ in the middle of the object is of order $\Delta v \aboutequal v_0~ a (N/R)$.  Evidently this $\Delta v$ is indeed negligible compared to $v_0$ provided $a\lessthanorabout R/N \aboutequal a^2/(R \phi_s)$.  That is, $\phi_s \lessthanorabout (a/R)$.  This restriction on $\phi_s$ thus becomes indefinitely strong as $R\goesto \infinity$.  Conversely, for $\phi_s$ much larger than this threshold, the screening effect of the stokeslets becomes strong, as we now discuss.

We consider the stokeslet concentration needed to block the external flow within some distance $\xi \muchlessthan R$ from the surface of the object.  For this we may consider stationary stokeslets distributed through a shell of finite thickness $d$ cloaking the object surface \cite{TWbook, Kirkwood-Riseman}. Thus the density $n$ of stokeslets is of order $N/(R^2 d)$.  Given a velocity $v_0$ on the outside of the shell, the flow diminishes with depth as the stokeslets exert a retarding force.  In steady state the drag force per unit area $6\pi ~\eta ~n a  v~ dx$ retarding a slab of width $dx$ is balanced by the difference of viscous stress across the slab: $\eta (dv/dx |_{x + dx} - dv/dx |_x) $, so that  $d^2v/dx^2 = (6 \pi n a) v(x)$.  Evidently the speed $v$ decreases exponentially with depth with a decay length $(6 \pi n a)^{-1/2}$ called the hydrodynamic screening length.  The stokeslets retard the external flow within a distance of the order of this decay length.

We now ask how many stokeslets per unit area are needed to confine the external flow to a given depth $\xi$.  Using the decay length as an estimate of $\xi$, we infer that $\xi$ is of order  $(R^2 d/(Na))^{1/2}$.  We next note that the thickness $d$ need not be greater than $\xi$ since at depth $\xi$ the flow is presumed sufficiently weak.  Thus taking $d \aboutequal \xi$, we obtain  $\xi \aboutequal R^2/(N a) \aboutequal a/\phi_s$.  Thus any $\xi$ however small may be obtained with arbitrarily dilute stokeslets (small $\phi_s$), provided the stokes radii of the stokeslets is sufficiently small.  In view of this scaling property, we define a   ``nominal penetration depth" $\xi_e$ used in the main text as $a/\phi_s$.

Section \ref{sec:discussion} compares this $\xi_e$ with the flow penetration observed in our stokeslet spheres. We find that $\xi_e$ is consistent with the observed flow near these spheres, even though the area fraction was not small.  As shown above, one could have attained the same degree of screening with a lower area fraction at the expense of making the stokeslets smaller and more numerous.

\section{Method for numerically determining total electric field $\vector E(\vector r)$}\label{sec:numEfield}

Here we outline a numerical method to determine the electric field $\vector E(\vector r)$ as modified by an insulating body of arbitrary shape.  As noted in Section \ref{sec:dpfield}, this field consists of the applied field $\vector E_0$ plus a depolarization field owing to depolarization charges on the surface, so as to remove any normally directed $\vector E$ at the surface.  This field depends only on the shape of the body and is independent of any explicit body charges. For a sphere it has the simple closed form of Eq. \eqref{eq:depfield}.  For general shapes, another method is needed.  Packaged programs to solve Laplace's equation for the potential with the needed surface boundary condition are readily available.  But for our discrete system it is convenient to use a similar discrete method to determine $\vector E$.  Accordingly, we assign a depolarization charge $Q_\alpha$ to each surface stokeslet, on the otherwise neutral body and adjust the $Q$'s so that 
\begin{equation} 
0 = \hat{\vector n}(\vector r^\alpha) \cdot \bigg (\vector E_0(\vector r^\alpha) + C' ~Q_\alpha \hat{\vector n}  
+ \sum_{\beta\notequal \alpha}  Q_\beta {\vector r^{\alpha} - \vector r^\beta \over |\vector r^\alpha - \vector r^\beta|^3}
\bigg ) 
\label{eq:EofQ}
\end{equation}
where again the term with $C'$ accounts for the self-field of charge $Q_\alpha$ at $\vector r^\alpha$. It is normally directed and proportional to the charge density at $\vector r^\alpha$, \ie the charge $Q_\alpha$ divided by the area per charge.
Once the charges $Q$ have been determined, we may use Eq. \eqref{eq:EofQ} to find $\vector E(\vector r)$ anywhere.  It is simply the quantity in parentheses with $\vector r^\alpha$ replaced by $\vector r$.  Since the needed $\vector r$'s are not at the $Q$ sites, we may sum over all $\beta$ and omit the $C'$ term. Preliminary results not reported here indicate that using Eq. \eqref{eq:EofQ} with stokeslet objects like those of Table \ref{nonuniformmobilitytable} is feasible.

\bibliography{StokesletElectrophoresis}

\end{document}